\documentclass[%twocolumn,doublespacing,superscriptaddress,
fleqn]{elsart}
\usepackage{amsmath,amssymb,dcolumn}

\def\be{\begin{equation}}
\def\ee{\end{equation}}
\def\bea{\begin{eqnarray}}
\def\eea{\end{eqnarray}}
\def\nn{\nonumber}

\begin{document}
\begin{frontmatter}
\begin{flushright}
arXiv: 0705.3871 [hep-th] \\
\textit{to appear in} Phys.Lett. B
\end{flushright}

\title{Anomalies and de Sitter radiation from the generic black holes in de Sitter spaces}
\author[1]{Qing-Quan Jiang},
\ead{jiangqingqua@126.com}
\author[2]{Shuang-Qing Wu},
\ead{sqwu@phy.ccnu.edu.cn}
\author[1]{Xu Cai}
%\ead{xcai@mail.ccnu.edu.cn}
\address[1]{Institute of Particle Physics, Central China Normal University,
Wuhan, Hubei 430079, People's Republic of China}
\address[2]{College of Physical Science and Technology, Central China Normal University, Wuhan,
Hubei 430079, People's Republic of China}
%\date{\today}

\begin{abstract}
Robinson-Wilczek's recent work shows that, the energy momentum tensor flux required to cancel
gravitational anomaly at the event horizon of a Schwarzschild-type black hole has an equivalent
form to that of a $(1+1)$-dimensional blackbody radiation at the Hawking temperature. Motivated
by their work, Hawking radiation from the cosmological horizons of the general Schwarzschild-de
Sitter and Kerr-de Sitter black holes, has been studied by the method of anomaly cancellation.
The result shows that the absorbing gauge current and energy momentum tensor fluxes required to
cancel gauge and gravitational anomalies at the cosmological horizon are precisely equal to those
of Hawking radiation from it. It should be emphasized that the effective field theory for generic
black holes in de Sitter spaces should be formulated within the region between the event horizon
(EH) and the cosmological horizon (CH), to integrate out the classically irrelevant ingoing modes
at the EH and the classically irrelevant outgoing modes at the CH, respectively.
\end{abstract}

\begin{keyword}
Hawking radiation \sep Anomaly \sep Cosmological horizon

\PACS 04.70.Dy \sep 04.62.+v \sep 03.65.Sq
\end{keyword}
\end{frontmatter}

\newpage

%%%%%%%%%%%%%%%%%%%%%%%%%%%%%%%%%%%%%%%%%%%%%%%%%%%%%%%%%%%%%
\section{Introduction}\label{intro}
%%%%%%%%%%%%%%%%%%%%%%%%%%%%%%%%%%%%%%%%%%%%%%%%%%%%%%%%%%%%%

Since Hawking radiation from black holes was first discovered by Stephen Hawking \cite{SWH1,SWH2}, many
derivations of Hawking radiation appeared in the literature (see, for example, \cite{PW,PW1,GH,CF}).
Recently, Robinson and Wilczek suggested that Hawking radiation can also be determined by anomaly
cancellation conditions and regularity requirement at the event horizon \cite{RW}. After a dimensional
reduction technique at the event horizon, quantum fields in the original higher dimensional theories
can be effectively treated as an infinite collection of two-dimensional quantum fields. In the effective
two-dimensional theory, the ingoing(outgoing) modes at the event horizon behave as the left(right)-moving
ones, respectively. Since the event horizon (EH) is a null hypersurface, the ingoing modes at the EH
that fall into the black hole would not classically affect the physics outside the black hole. Quantum
mechanically, however, their quantum contribution to the physics outside the EH should be taken into
account. If the effective field theory is formulated outside the EH to exclude the classically irrelevant
ingoing modes at the EH, it becomes chiral there and contains a gravitational anomaly, which takes the
form of the nonconservation of the energy momentum tensor. To cancel gravitational anomaly at the EH
and to restore general coordinate covariance at the quantum level, one must introduce a compensate
energy momentum tensor flux, which is shown to be exactly equal to that of Hawking radiation. This
is the basic idea of Robinson-Wilczek's to derive of Hawking radiation via the anomalous point of
view. The method was then soon generalized to the cases of charged \cite{IUW1} and rotating black
holes \cite{MS,IUW2} by considering gauge and gravitational anomalies at the horizon, and further
applied to other cases \cite{VD,MRS,XC,QSX,QS,STH}.

Till now, a common feature that shares in these observations is that both gauge and gravitational
anomalies take place at the event horizon of black holes with or without a cosmological constant.
In Ref. \cite{XC}, although the authors attempted to extend the Robinson-Wilczek's work to general
Kerr-de Sitter black holes in $D$ dimensions, they only studied Hawking radiation from the EH of
the black hole. The situation that takes place at the cosmological horizon (CH), however, had not
been addressed in detail there. In fact, it was demonstrated \cite{GH} that particles can also be
created at the CH with a thermal spectrum. Although it is of no much practical significance in
astrophysics to enclose the de Sitter thermal radiation of the CH because its temperature carried
by the radiation many be very small, researches on black holes in de Sitter spaces become more and
more important at least due to the following two reasons: (1) The recent observed accelerating
expansion of our universe indicates the cosmological constant might be a positive one \cite{ASJ};
and (2) Conjecture about de Sitter/conformal field theory correspondence \cite{AD}. Thus it many
be of special interest to carefully investigate Hawking radiation via anomaly cancellation at the
CH of black holes with a postive cosmological constant.

In Ref. \cite{XC}, the effective field theory in the two-dimensional reduction of general Kerr-de
Sitter black holes is formulated outside the EH to integrate out the classically irrelevant ingoing
modes at the EH. In fact, in the case of general black holes with a repulse cosmological constant,
such as the generic Schwarzschild-de Sitter and Kerr-de Sitter black holes, there exist event horizon
and cosmological horizon for an observer moving along the world line of constant $r$ between both
horizons \cite{GH}. In such cases, the effective field theory that only describes observable physics
should be formulated within the region between the EH and the CH to exclude the classically irrelevant
ingoing modes at the EH and the classically irrelevant outgoing modes at the CH. Thus, gauge and
gravitational anomalies arise both at the EH and at the CH.

In this Letter, our main motivation is to study Hawking radiation from the CH via anomaly cancellation.
To simplify our discussions, for the moment when we study Hawking radiation from the CH, we can regard
that the gauge and gravitational anomalies taken place in the vicinity of the CH are only due to exclude
the classically irrelevant outgoing modes at the CH, and ignore the quantum contribution of the omitted
ingoing modes at the EH although they should be incorporated into the effective theory formulated at
the EH, that is to say, the EH and the CH are considered as two independent physical systems with their
probably interactions being overlooked. A similar recipe to deal with Hawking radiation via tunnelling
from the CH of black holes in de Sitter spaces had already been successfully written out in \cite{AJ}.

At the CH, the outgoing modes that fall out of the CH would not classically fall back since the CH is
also a null hypersurface. Quantum mechanically, however, the quantum contribution of the classically
irrelevant outgoing modes to the physics inside the CH should be taken into account. If the effective
field theory is formulated inside the CH to integrate out the classically irrelevant outgoing modes
at the CH, it become chiral there and contains gauge and gravitational anomalies with respect to gauge
and general coordinate symmetries. To cancel these anomalies and to restore gauge and general coordinate
covariance at the quantum level in the effective field theory, each partial wave of the scalar field
must be in a state with a gauge current flux and an energy momentum tensor flux. Our result shows that
the absorbing gauge current and energy momentum tensor fluxes, required to cancel gauge and gravitational
anomalies at the CH, are precisely equal to those of Hawking radiation from the CH.

Our Letter is outlined as follows. By extending the Robinson-Wilczek's work \cite{RW} that Hawking
radiation can be determined by anomaly cancellation conditions and regularity requirement at the EH,
we investigate, in Sec. \ref{sds}, Hawking radiation from the CH of a generic Schwarzschild-de Sitter
black hole. Sec. \ref{kds} is devoted to investigating Hawking radiation from the CH of the general
Kerr-de Sitter black hole via cancellation of gauge and gravitational anomalies. In both cases, we
adopt, for simplicity, the viewpoint that gauge and gravitational anomalies taken place in the vicinity
of the CH are due to exclude the classically irrelevant outgoing modes at the CH, and neglect quantum
contribution of the omitted ingoing modes at the EH. Sec. \ref{dc} ends up with
our conclusions.

%%%%%%%%%%%%%%%%%%%%%%%%%%%%%%%%%%%%%%%%%%%%%%%%%%%%%%%%%%%%%
\section{Hawking radiation from the CH of the generic Schwarzschild-de Sitter black holes}\label{sds}
%%%%%%%%%%%%%%%%%%%%%%%%%%%%%%%%%%%%%%%%%%%%%%%%%%%%%%%%%%%%%

The metric of a general Schwarzschild-de Sitter black hole can be expressed as
\be
ds^2 = f(r)dt^2 -\frac{1}{f(r)}dr^2 -r^2d\Omega_{n}^2 \, ,\label{sdse}
\ee
where
\be
f(r) = 1 -\frac{\omega_n M}{r^{n-1}} -\frac{r^2}{l^2} \, , \qquad
\omega_n = \frac{16\pi G}{n V_{n}} \, ,
\ee
in which $M$ is the mass of the black hole, $l$ is the curvature radius of de Sitter space, and $V_n$
denotes the volume of a unit $n$-sphere $d\Omega_n^2$. (Units $G_{n+2} = c = \hbar = 1$ are adopted
throughout this article). When $M = 0$, the solution (\ref{sdse}) reduces to the pure de Sitter space
with a cosmological horizon $r = l$ which may be very large according to the existing knowledge of
the cosmological constant. For the general case in higher dimensions ($n \geq 2$), the horizons are
determined by the equation $f(r) = 0$ which is of ($n+1$)-order so it in general has ($n+1$)-roots.
Typically for $n \geq 2$, there will be two positive (real) roots of $f(r) = 0$, with the outermost
root ($r_c$) representing a cosmological horizon, and the remaining one ($r_h$) describing the black
hole horizon. The explicit forms of these solutions are not needed for our discussions made here and
are not illuminating though a detailed analysis may be of some special interest.

As mentioned above, all existing investigations \cite{RW,IUW1,MS,IUW2,VD,MRS,XC,QSX,QS,STH} on Hawking
radiation via the anomaly cancellation method are based upon the effective theory established in the
vicinity of the black hole event horizon (EH). In other words, the effective field theory is always
formulated outside the EH to exclude the classically irrelevant ingoing modes at the EH. In fact, in
the case of black holes in de Sitter spaces, since there exist two horizons: the EH and the CH, the
effective field theory that only describes observable physics should be formulated within the region
between the EH and the CH, to integrate out the classically irrelevant ingoing modes at the EH and
the classically irrelevant outgoing modes at the CH, respectively. For the sake of simplicity, when
studying Hawking radiation from the EH, one can only treat gauge and gravitational anomalies taken
place in the vicinity of the EH as arising from excluding the classically irrelevant ingoing modes
at the EH, and overlook the quantum contribution of the omitted outgoing modes at the CH although
they should be incorporated into the effective field theory in the vicinity of the CH. Similar measures
can be taken to tackle with the case when one considers the de Sitter radiation. Subsequently, one can
apply the same procedure as did in Refs. \cite{RW,IUW1,MS,IUW2,VD,MRS,XC,QSX,QS,STH} to derive Hawking
radiation from the EH via the anomalous point of view. However, since our main propose is to study
Hawking radiation from the CH of the generic black holes in de Sitter spaces, we can, similarly,
regard the gauge and gravitational anomalies taken place in the vicinity of the CH are only due to
integrate out the classically irrelevant outgoing modes at the CH, and disregard the quantum effect
of the omitted ingoing modes at the EH although they should be incorporated into the effective field
theory in the vicinity of the EH.

Now we concentrate on studying Hawking radiation from the cosmological horizon of a generic Schwarzschild-de
Sitter black hole via anomaly cancellation. Near the CH, if one introduces the tortoise coordinate
transformation defined by $dr/dr_* = f(r)$, and performs the partial wave decomposition in terms of
spherical harmonics, the effective radial potential for partial wave modes of the scalar field vanishes
exponentially fast \cite{RW,IUW1}. Thus physics near the CH can be described by an infinite collection
of $(1+1)$-dimensional fields, each partial wave propagating in a spacetime with the effective metric
given by
\be
ds^2 = f(r)dt^2 -\frac{1}{f(r)}dr^2 \, .
\ee
In addition, the background also includes a dilaton field, whose contribution can be dropped due to
the static background \cite{IUW1}. In the two-dimensional reduction, gravitational anomaly taken place
at the CH is due to exclude the classically irrelevant outgoing modes at the CH. For the left-handed
field (ingoing modes), the consistent gravitational anomaly takes the form as
\be
\nabla_\mu T_\nu^\mu = -\frac{1}{\sqrt{-g}}\partial_\mu \mathcal{N_\nu^\mu} \, ,
\ee
where
\be
\mathcal{N_\nu^\mu} = \frac{1}{96\pi}\epsilon^{\beta\mu}\partial_\alpha \Gamma_{\nu \beta}^\alpha \, .
\ee

As we are simply considering the de Sitter radiation for which the effective field theory should be
formulated in the region between the EH and the CH, we can only focus on the physics taken place near
the CH and ignore the effect of the EH. The total energy momentum tensor is composed of a sum from
two regions: the near-horizon region ($r_c -\epsilon \leq r \leq r_c$) and the other region ($r_h \ll
r \leq r_c -\epsilon$), that is, $T_\nu^\mu = T_{\nu(o)}^\mu\Theta_- +T_{\nu(C)}^\mu C$, where $\Theta_-
= \Theta(r_c -r -\epsilon)$ and $C = 1 -\Theta_-$ are, respectively, a scalar step and top hat function.
Near the CH ($r_c -\epsilon \leq r \leq r_c$), gravitational anomaly taken place in the effective field
theory gives an important constraint on the energy momentum tensor as
\be
\partial_r T_{t(C)}^r = -\partial_r \mathcal{N}_t^r(r) \, , \qquad
\mathcal{N}_t^r(r) = \frac{1}{192\pi}\big(f'^2 +ff''\big) \, . \label{N}
\ee
In the other region, there is no anomaly, the energy momentum tensor in this region satisfies the
conservation equation $\partial_rT_{t(o)}^r = 0$. In the classical theory, general coordinate covariance
of the classical action demands $-\delta W = \int d^2x\sqrt{-g}\lambda^\nu \nabla_\mu T_\nu^\mu = 0$,
where $\lambda^\nu$ is a variation parameter. In the case of de Sitter radiation at the CH, the effective
theory excludes the classically irrelevant outgoing modes at the CH, but should include the quantum
contribution of the omitted ingoing modes at the EH, whose effect on the CH, however, had not been
considered in the discussions below for the sake of simplicity. The variance of the effective action
under the general coordinate transformation can be written as
\bea
-\delta W &=& \int dtdr \lambda^t \nabla_\mu\big(T_{t(o)}^\mu \Theta_- +T_{t(C)}^\mu C\big) \nn \\
&=& \int dtdr \lambda^t\Big[\big(T_{t(C)}^r -T_{t(o)}^r +\mathcal{N}_t^r\big)
 \delta (r -r_c +\epsilon) -\partial_r\big(C\mathcal{N}_t^r\big)\Big] \, . \label{W1}
\eea
In Eq. (\ref{W1}), the last term should be cancelled by the quantum effect of the classically irrelevant
outgoing modes at the CH, whose contribution to the total energy momentum tensor is $C\mathcal{N}_t^r$.
To restore general coordinate covariance at the quantum level, the coefficient of the delta function
should vanish, which means
\be
a_o = a_c +\mathcal{N}_t^r \, ,
\ee
where
\bea
a_o &=& T_{t(o)}^r \, , \nn \\
a_c &=& T_{t(C)}^r +\int_{r_c}^rdr\partial_r\mathcal{N}_t^r \, ,
\eea
are, respectively, the flux of the energy momentum tensor observed by an observer who inhabits in the
region between the EH and the CH, and the one at the CH. To ensure the regularity for the energy momentum
tensor, we impose a vanishing condition for the covariant energy momentum tensor at the CH. Since the
covariant energy momentum tensor is related to the consistent one by
\be
\widetilde{T}_t^r = T_t^r -\frac{1}{192\pi}\big(ff'' -2f'^2\big) \, , \label{T2}
\ee
that condition yields
\be
a_c = -\frac{\kappa_c^2}{24\pi} = -2\mathcal{N}_t^r(r_c) \, ,
\ee
where $\kappa_c = -\partial_r f(r)/2|_{r = r_c}$ is the surface gravity at the CH. The total energy
momentum tensor flux is then given by
\be
a_o = -\mathcal{N}_t^r(r_c) = -\frac{\pi}{12}T_c^2 \, , \label{ao}
\ee
where $T_c = \kappa_c/(2\pi)$ is the Hawking temperature at the CH of the black hole.

In Eq. (\ref{ao}), the negative sign $(-)$ demonstrates that in the effective field theory the energy
momentum tensor flux must be absorbed at the CH in order to ensure the general coordinate covariance
at the quantum level. In contrary, one must take the positive sign $(+)$ for the compensating energy
momentum tensor flux at the EH in order to cancel gravitational anomaly at the EH
\cite{RW,IUW1,MS,IUW2,VD,MRS,XC,QSX,QS,STH}.

To cancel gravitational anomaly at the CH and to restore general coordinate covariance at the quantum
level, the energy momentum tensor flux radiated into the black hole from the CH must be given by Eq.
(\ref{ao}). In fact, this absorbing energy momentum tensor flux has an equivalent form as that of Hawking
radiation from the CH of the black hole. At the CH, since blackbody radiation is moving along the negative
$r$ direction, its Planckian distribution with the Hawking temperature $T_c$ is written as $\mathcal{N}
(\omega) = -1/[\exp(\frac{\omega}{T_c}) +1]$ for fermions. [By contrast, the sign of the Planckian
distribution of blackbody radiation moving in the positive $r$ direction is positive at the EH.]
With this distribution, the energy momentum tensor flux reads
\be
F_c = \int_0^\infty \frac{\omega}{\pi}\mathcal{N}(\omega)d\omega
= -\frac{\pi}{12}T_c^2 \, . \label{Fc}
\ee
Compare Eqs. (\ref{ao}) with (\ref{Fc}), we find that the absorbing energy momentum tensor flux, required
to cancel gravitational anomaly at the CH and to restore general coordinate covariance at the quantum
level in the effective field theory, is exactly equal to that of Hawking radiation from the CH. This
result shows that the flux of Hawking radiation from the CH can be determined by anomaly cancellation
conditions and regularity requirement at the CH.

In the following section, we will further extend the Robinson-Wilczek's work to the case of a generic
Kerr-de Sitter black hole. In Ref. \cite{XC}, Hawking radiation from the event horizon of a general
Kerr-de Sitter black hole has been studied by the anomalous point of view, where the effective field
theory is formulated outside the EH to exclude the classically irrelevant ingoing modes at the EH. In
fact, in the case of rotating black holes with a repulse cosmological constant, the effective field
theory that only describes observable physics should also be formulated within the region between the
EH and the CH to integrate out the classically irrelevant ingoing modes at the EH and the classically
irrelevant outgoing modes at the CH, respectively. Thus gauge and gravitational anomalies can take place
at both the EH and the CH. In what follows, we shall only focus on studying Hawking radiation from the
CH. To simplify our discussion, we will also disregard the effect of the EH when we consider the de
Sitter radiation.

%%%%%%%%%%%%%%%%%%%%%%%%%%%%%%%%%%%%%%%%%%%%%%%%%%%%%%%%%%%%%
\section{Hawking radiation from the CH of the general Kerr-de Sitter black hole}\label{kds}
%%%%%%%%%%%%%%%%%%%%%%%%%%%%%%%%%%%%%%%%%%%%%%%%%%%%%%%%%%%%%

The metric of a general Kerr-de Sitter black hole in $D$ dimension takes the form in a Boyer-Lindquist
coordinate system as \cite{GHDC}
\bea
ds^2 &=& W(1 -\lambda r^2)dt^2 -\frac{2M}{VF}
 \Big(Wdt -\sum_{i = 1}^N \frac{a_i\mu_i^2}{1 +\lambda a_i^2}d\varphi_i\Big)^2 \nn \\
&& -\sum_{i = 1}^N\frac{r^2 +a_i^2}{1 +\lambda a_i^2}\mu_i^2 d\varphi_i^2
  -\frac{VF}{V-2M}dr^2-\sum_{i = 1}^{N+\epsilon}
 \frac{r^2 +a_i^2}{1+ \lambda a_i^2}d\mu_i^2 \nn \\
&&\quad -\frac{\lambda}{W(1 -\lambda r^2)} \Big(\sum_{i = 1}^{N +\epsilon}
 \frac{r^2 +a_i^2}{1 +\lambda a_i^2}\mu_i d\mu_i\Big)^2 \, ,
\eea
where $\epsilon = 0$, $1$ correspond to odd and even dimensions, respectively, and
\bea
&W& = \sum_{i = 1}^{N +\epsilon}\frac{\mu_i^2}{1 +\lambda a_i^2} \, , \qquad
F = \frac{1}{1 -\lambda r^2}\sum_{i = 1}^{N +\epsilon}\frac{r^2\mu_i^2}{r^2 +a_i^2} \, , \nn\\
&V& = r^{\epsilon -2}(1 -\lambda r^2)\prod_{i = 1}^{N}(r^2 +a_i^2) \, ,
\eea
where $N$ is the integral part of $(D-1)/2$, $\mu_i$ should satisfy the following constraint
$\sum_{i = 1}^{N +\epsilon}\mu_i^2 = 1$, and we assume the cosmological constant $\lambda > 0$.
There are $N$ independent rotation parameters $a_i$ in orthogonal spatial $2$-planes in general.
Near the CH, introducing the tortoise coordinate transformation $dr/dr_* = f(r)$, where
$f(r) \approx -2\kappa_c(r -r_c)$ in which
\be
\kappa_c = \frac{1}{2}\big(1 -\lambda r_c^2\big)\frac{V'(r_c)}{V(r_c)} \, ,
\ee
is the surface gravity at the CH of the black hole, and further performing the partial wave decomposition
$\phi = \sum_{m_i}\phi_{m_1,\dots, m_N}(t,r)Y_{m_1,\dots, m_N}(\mu_i, \varphi_i)$ \cite{XC}, physics near
the CH can be described by using an infinite collection of $(1+1)$ dimensional fields, each propagating
in the backgrounds of the effective metric $g_{\mu\nu}$ and $U(1)$ gauge fields $A_\mu^i$ as follows
\bea
&g_{tt}& = f(r) \, , \qquad g_{rr} = -f(r)^{-1} \, , \qquad g_{tr} = 0 \, , \nn\\
&A_t^i& = -\frac{a_i(1 -\lambda r^2)}{r^2 +a_i^2} \, , \qquad A_r^i = 0 \, . \label{g}
\eea

The $U(1)$ charges of the two-dimensional field are the azimuthal quantum numbers along $\varphi_i$
direction $m_i$. In addition to general coordinate symmetry, the effective two-dimensional theory
contains $N$ $U(1)$ gauge symmetries. In order to investigate the de Sitter radiation, the effective
field theory should be formulated in the region between the EH and the CH to cancel the $U(1)$ gauge
and gravitational anomalies. As before, we shall only consider the physics near the CH and overlook
the effect on the CH coming from the EH. In the near-horizon region $r_c-\epsilon \leq r \leq r_c$,
the effective theory is chiral and contains gauge and gravitational anomalies. For the left-handed
field (ingoing modes), the consistent $U(1)$ gauge anomaly equation reads off
\be
\nabla_\mu J_i^u = -\frac{m_i}{4\pi}\epsilon^{\mu\nu}\partial_\mu \mathcal{A}_\nu \, ,
\ee
where $\mathcal{A}_\nu = m_iA_\nu^i$ is the sum of the $N$ U(1) gauge fields. Since the anomaly is
purely timelike, the anomaly equation for each $U(1)$ gauge current near the CH is given by $\partial_r
J_{i(C)}^r = -{m_i}\partial_r\mathcal{A}_t/({4\pi})$. In the other region, no $U(1)$ gauge anomaly
takes place, each current is conserved and satisfies $\partial_r J_{i(o)}^r = 0$. When omitting the
classically irrelevant outgoing modes at the CH, the total $U(1)$ gauge current is consisted of a sum
from two regions $J_i^\mu = J_{i(o)}^\mu \Theta_- +J_{i(C)}^\mu C$. Under the gauge transformation,
the effective action changes as
\bea
-\delta W &=& \int dtdr \lambda \nabla_\mu\big(J_{i(o)}^\mu \Theta_- +J_{i(C)}^\mu C\big) \nn\\
&=&\int dtdr \lambda \Big[\big(J_{i(C)}^r -J_{i(o)}^r +\frac{m_i}{4\pi}
 \mathcal{A}_t\big)\delta (r -r_c +\epsilon) -\partial_r\big(\frac{m_i}{4\pi}
 \mathcal{A}_t C\big)\Big] \, , \qquad \label{W2}
\eea
where $\lambda$ is a gauge parameter. In Eq. (\ref{W2}), we have omitted the classically irrelevant
outgoing modes at the CH, whose contribution to the total gauge current is ${m_i}\mathcal{A}_t C/({4\pi})$.
The second term should be cancelled by their quantum contribution. Since the underlying theory must be
gauge invariant, the coefficient of the delta function should vanish, which says that
\be
d_{i(o)} = d_{i(c)} +\frac{m_i}{4\pi}\mathcal{A}_t(r_c) \, ,
\ee
where $d_{i(o)} = J_{i(o)}^r$ is the gauge current flux observed by an observer who lives in the region
between the EH and the CH, and
\be
d_{i(c)} = J_{i(C)}^r +\frac{m_i}{4\pi}\int_{r_c}^r dr \partial_r \mathcal{A}_t \, ,
\ee
is the one at the CH. In order to fix the current flux, we impose a constraint that the covariant current,
which is related to the consistent one by
\be
\widetilde{J}_i^r = J_i^r -\frac{m_i}{4\pi}\mathcal{A}_t C \, ,
\ee
vanishes at the CH. Using this condition, one can easily determine the value of the $U(1)$ gauge current
flux to be
\be
d_{i(o)} = \frac{m_i}{2\pi}\mathcal{A}_t(r_c) = -\frac{m_i}{2\pi}\sum_{j = 1}^N m_j
\frac{a_j(1 -\lambda r_c^2)}{r_c^2 +a_j^2} \, . \label{dio}
\ee
This flux corresponds to the angular momentum flux of Hawking radiation from the CH of the black hole.
The negative sign reflects that the $U(1)$ gauge current flux is radiated into the black hole from the
CH.

In addition to the $U(1)$ gauge symmetries, there is also the general coordinate symmetry in the effective
two-dimensional theory. When excluding the horizon-skimming modes at the horizons, the effective field
theory contains both gauge and gravitational anomalies. As before, we shall only deal with Hawking radiation
via anomaly cancellation from the CH, and disregard the effect of the EH, for the simplicity. In the region
near the CH, $U(1)$ gauge and gravitational anomalies constraint the energy momentum tensor by
\be
\partial_r T_{t(C)}^r = \mathcal{J}_{(C)}^r\partial_r \mathcal{A}_t +\mathcal{A}_t \partial_r
\mathcal{J}_{(C)}^r -\partial_r\mathcal{N}_t^r \, . \label{T1}
\ee
In Eq. (\ref{T1}), $\mathcal{N}_t^r$ takes the same form as before with $f(r)$ now given by Eq. (\ref{g}).
Also, we have $\mathcal{J}^r \equiv J_i^r/m_i = J_j^r/m_j$. In the other region, there is no anomaly, and
the energy momentum tensor is conserved. In a background with $U(1)$ gauge fields, the conservation equation
for the energy momentum tensor is modified as $\partial_rT_{t(o)}^r = \mathcal{J}_{(o)}^r\partial_r
\mathcal{A}_t$, in which
\be
\mathcal{J}_{(o)}^r = \frac{1}{2\pi}\mathcal{A}_t(r_c) = -\frac{1}{2\pi}\sum_{j = 1}^N m_j
\frac{a_j(1 -\lambda r_c^2)}{r_c^2 +a_j^2}\equiv d_o \, .
\ee

In the simplest case we are considering here, the total energy momentum tensor combines contribution from
two regions, that is, $T_\nu^\mu = T_{\nu(o)}^\mu\Theta_- +T_{\nu(C)}^\mu C$. Under the general coordinate
transformation, the effective action (omitting the classically irrelevant outgoing modes at the CH) changes
as
\bea
-\delta W &=& \int dtdr \lambda^t\nabla_\mu\big(T_{t(o)}^\mu \Theta_- +T_{t(C)}^\mu C\big) \nn\\
 &=& \int dtdr \lambda^t\Big[d_o \partial_r \mathcal{A}_t -\partial_r
 \big(\frac{1}{4\pi}\mathcal{A}_t^2 +\mathcal{N}_t^r\big)C \nn \\
 && +\big(T_{t(C)}^r -T_{t(o)}^r +\frac{1}{4\pi}\mathcal{A}_t^2
 +\mathcal{N}_t^r\big)\delta (r -r_c +\epsilon)\Big] \, . \label{W3}
\eea
In Eq. (\ref{W3}), the first term is the classical effect of the background gauge field for constant
current flow. The second term should be cancelled by the quantum effect of the classically irrelevant
outgoing modes at the CH, whose contribution to the total energy momentum tensor is $[\mathcal{A}_t^2/(4\pi)
+\mathcal{N}_t^r]C$. To restore general coordinate covariance at the quantum level, the coefficient
of the delta function should vanish, thus we have
\be
f_o = f_c -\frac{1}{4\pi}\mathcal{A}_t^2(r_c) +\mathcal{N}_t^r(r_c) \, , \label{fo}
\ee
where $f_o = T_{t(o)}^r -d_o\mathcal{A}_t$ is the energy flow observed by an observer who lives in
the region between the EH and the CH, and
\be
f_c = T_{t(C)}^r -\int_{r_c}^r dr \partial_r \big(d_o\mathcal{A}_t
 -\frac{1}{4\pi}\mathcal{A}_t^2 -\mathcal{N}_t^r\big) \, ,
\ee
is the energy momentum tensor flux at the CH. Here we have used ${J}_{(C)}^r = d_o -\mathcal{A}_t/({4\pi})$.
Similarly, we impose a constraint that the covariant energy momentum tensor vanishes at the CH, so that
the total energy momentum tensor flux is given by
\bea
f_o &=& -\mathcal{N}_t^r(r_c) -\frac{1}{4\pi}\mathcal{A}_t^2(r_c) \nn \\
 &=& -\frac{1}{4\pi}\Big(\sum_{i = 1}^N m_i \frac{a_i(1 -\lambda r_c^2)}{r_c^2 +a_i^2}\Big)
 -\frac{\pi}{12}T_c^2 \, , \label{fo1}
\eea
where $T_c = \kappa_c/(2\pi)$ is the Hawking temperature at the CH of the black hole. As mentioned before,
the negative sign means that the energy momentum tensor flux is radiated into the black hole from the CH.

In fact, these absorbing gauge current and energy momentum tensor fluxes in Eqs. (\ref{dio}) and (\ref{fo1}),
which are required to restore gauge invariance and general coordinate covariance at the quantum level, have
equivalent forms to those of Hawking radiation from the CH of the black hole. For the case of fermions, the
Hawking distribution at the CH takes the form $\mathcal{N}_{\pm m}(\omega) = -1/[\exp(\frac{\omega\pm m_i
A_t^i(r_c)}{T_c}) +1]$ (Note: the negative sign means that de Sitter radiation is radiated into the black
hole from the CH). Integrating with respect to this distribution, the angular momentum and energy momentum
tensor fluxes at the CH can be shown to take the same forms as Eqs. (\ref{dio}) and (\ref{fo1}),
respectively. This indicates that de Sitter radiation can also be determined by the method of cancellation
of anomaly.

%%%%%%%%%%%%%%%%%%%%%%%%%%%%%%%%%%%%%%%%%%%%%%%%%%%%%%%%%%%%%
\section{Conclusions}\label{dc}
%%%%%%%%%%%%%%%%%%%%%%%%%%%%%%%%%%%%%%%%%%%%%%%%%%%%%%%%%%%%%

Motivated by Robinson-Wilczek's recent work \cite{RW}, we have studied Hawking radiation from the
cosmological horizon of the general Schwarzschild-de Sitter and general Kerr-de Sitter black holes
via the anomalous point of view. The result shows that the absorbing gauge current and energy momentum
tensor fluxes, required to cancel gauge and gravitational anomalies at the CH and to restore gauge
invariance and general coordinate covariance at the quantum level, are exactly equal to those of
Hawking radiation from the CH. This is very similar to the case taken place at the EH, however,
several points deserve to be emphasized:

i) Gauge and gravitational anomalies taken place at the CH are due to exclude the classically irrelevant
outgoing modes at the CH.

ii) For general black holes in de Sitter spaces, the effective field theory that only describes observable
physics should be formulated within the region between the EH and the CH to integrate out the classically
irrelevant ingoing modes at the EH and the classically irrelevant outgoing modes at the CH, respectively.

iii) When dealing with Hawking radiation from the CH, we have taken the simplest case that gauge and
gravitational anomalies taken place in the effective theory are due to exclude the classically irrelevant
outgoing modes at the CH, and disregarded the effective field theory that contains the quantum contribution
of the omitted ingoing modes at the EH. In other words, we have assumed that the EH and the CH behave
like two independent systems, and overlooked the effect coming from the EH when we consider the de Sitter
radiation from the CH.

\section*{Acknowledgments}

This work was partially supported by the Natural Science Foundation of China under Grant Nos. 10675051,
10635020, 70571027, 70401020 and a grant by M.O.E under Grant No. 306022.

\end{document}